\documentclass[aip,reprint]{revtex4-1}

\draft 

\begin{document}


\title{Conformal supermultiplets without superpartners} 





\author{Peter D. Jarvis}
\affiliation{
School of Mathematics and Physics, University of Tasmania, Australia\\
\texttt{\small Peter.Jarvis@utas.edu.au}}
\date{June 2011}

\begin{abstract}
We consider polynomial deformations of Lie superalgebras and their representations.
For the class $A(n\!-\!1,0)\cong sl(n/1)$, we identify 
families of superalgebras of quadratic and cubic type, consistent with Jacobi identities.
For such deformed superalgebras we point out the possibility of \emph{zero step} supermultiplets, carried on a single, irreducible representation of the even (Lie) subalgebra. For the conformal group $SU(2,2)$ in $1\!+\!3$-dimensional spacetime, such  irreducible (unitary) representations correspond to standard conformal fields $(j_1,j_2;d)$, where $(j_1,j_2)$ is the spin and $d$ the conformal dimension; in the massless class $j_1j_2=0$, and $d=j_1+j_2+1$. We show that these repesentations are zero step supermultiplets for the superalgebra $SU_{(2)}(2,2/1)$, the quadratic deformation of conformal supersymmetry $SU(2,2/1)$. We propose to elevate $SU_{(2)}(2,2/1)$ to a symmetry of the $S$-matrix. Under this scenario, low-energy standard model matter fields (leptons, quarks, Higgs scalars and gauge fields) descended from such conformal supermultiplets are not accompanied by superpartners.
\end{abstract}

\pacs{}

\maketitle 

\section{Introduction}


%
%

%

The concept of supersymmetry as an extended relativistic invariance
principle for the fundamental particles and their interactions confronts
severe experimental constraints, as bounds on masses of the 
known particles' predicted superpartners are pushed upwards. In this paper we
point out that for certain natural extensions of spacetime
supersymmetries, \emph{there are representations without any superpartners}.

As is well known, supersymmetry in nature is handled at the technical
level by Lie superalgebras, generalisations of Lie algebras whose role
(especially in their integrated form as Lie groups) in the physical and
mathematical aspects of relativistic field theory is well known. Below,
we motivate the introduction of 
`deformations' of such superalgebras. Our main technical
result is the following: for a class of quadratic polynomial deformations of the non-extended
spacetime conformal superalgebra in 1+3 dimensions, there exist certain
unitary irreducible supermultiplets carried on only one module of the even
subalgebra: that is, which are comprised of a single, standard (massless)
conformal field of fixed spin, thus obviating the need for superpartner
fields.

We wish to generalise the Lie superalgebras which are relevant for
spacetime supersymmetries. In order to have a calculable test case for
our assertions about the remarkable possibilities for representations,
in the present work we eschew Poincar\'{e} supersymmetry 
and more exotic possibilities like $N$-extended supersymmetries or higher-dimensional, 
or string- or membrane- derived forms, in favour of
the algebraically more tractable conformal superalgebra, or
equivalently at the level of complex algebras, we examine the general
class $A(n-1,0) \cong sl(n/1)$ in the Kac-Dynkin classification \cite{Kac1977lsa} (specializing below to
the real $n=4$ form $su(2,2/1)$).

\section{Polynomial $gl(n)$ superalgebras}

The method consists simply in postulating generalised structure constants
for the bracket relations of the generators of these algebras, which give
the standard relations, plus (including deformation parameters), terms `nonlinear' in the
generators
(quadratic and higher polynomials, of degree $p \ge 2$). Such algebras are of course no longer of
Lie type,
but at the least we can demand that natural generalisations of the Jacobi
identities hold
(so that in realisations of the relations by linear operators, the
brackets are simply implemented by
the usual commutators and anticommutators).

In the simplest presentation, the superalgebra in question has even (Lie) part 
$gl(n)$. A basis is given by the Gel'fand generators $L^a{}_b$,
$a,b = 1,2,\cdots,n $, which fulfil the standard commutation relations 
\begin{equation}
{[}L^a{}_b,L^c{}_d{]} = \delta^c{}_b L^a{}_d - \delta^a{}_d L^c{}_b.
\label{eq:GLnGelfandBasis}
\end{equation}
The odd generators are $\overline{Q}{}^b$, $Q_a$, $a,b = 1,2,\cdots, n$, 
and transform as vector, and adjoint vector, with respect to the $L^a{}_b$:
\begin{equation}
{[}L^a{}_b,\overline{Q}{}^c{]} = \delta^c{}_b \overline{Q}{}^a, \qquad {[}L^a{}_b,Q_c{]} = - \delta^a{}_c Q_b.
\end{equation}
The two sets of odd generators are assumed to anticommute amongst themselves,
that is, ${\{} Q_a,Q_b {\}} = 0$ $={\{}\overline{Q}{}^a,\overline{Q}{}^b{\}}$. 
For the class of deformations which we wish to consider, these relations are unaltered; rather, we examine the rule for the anticommutator ${\{}\overline{Q}{}^a,{Q}_b{\}}$ with additional 
nonlinear terms. We state the following result, which for convenience
includes the standard Lie superalgebra case (degree $p=1$): \\[.2cm]
\emph{Theorem. Polynomial $gl(n)$ superalgebras}.\hfill \\ 
The anticommutation relations amongst odd generators defined by the following relations,
\begin{eqnarray}
p=1:\qquad {\{}\overline{Q}{}^a,{Q}_b{\}} &=& \lambda {\mathcal P}_{(1)}(L; \texttt{c})^a{}_b, \nonumber \\
p=2:\qquad{\{}\overline{Q}{}^a,{Q}_b{\}} &=& \lambda {\mathcal P}_{(2)}(L; \alpha,\texttt{c})^a{}_b, \nonumber \\
p=3:\qquad{\{}\overline{Q}{}^a,{Q}_b{\}} &=& \lambda {\mathcal P}_{(3)}(L; \alpha,\beta,\texttt{c})^a{}_b, 
\label{eq:CurlyPdefs}
\end{eqnarray}
satisfy the Jacobi identities and define (for degree $p \ge 2$) polynomial super-$gl(n)$ algebras. Here $\lambda$, $\alpha$, $\beta$ and \texttt{c} are complex parameters appropriate to each case.
The $p=2,3$ quadratic and cubic polynomials are given in Table \ref{tab:Polys}. There is no $p=4$ solution. For $p=1$ we have
${\mathcal P}_{(1)}{}^a{}_b =  L^a{}_b - \delta^a{}_b\langle L \rangle + \delta^a{}_b \texttt{c}$, which is equivalent to the Lie superalgebra
$sl(n/1)$. \hfill $\diamondsuit$
\\[.2cm]

In the above relations and in Table \ref{tab:Polys}, we use the following notation: $(L^n)^a{}_b = \sum_c (L^{n\!-\!1})^a{}_c L^c{}_b$, $(L^0)^a{}_b=\delta^a{}_b$,
and $\langle L^n \rangle = \sum_c (L^n)^c{}_c$ are the standard Casimir operators.
We denote the deformed superalgebras defined via the above relations by $sl_{(p)}(n/1)$, by analogy with the ordinary Lie superalgebra case. The explicit forms of the anticommutation relations embodied in the ${\mathcal P}_{(p)}(L)^a{}_b$ is best extracted from the tabulated coefficients (Table \ref{tab:Polys}); in the quadratic case: \hfill \mbox{}
\begin{widetext}
$$\mbox{for example for $p=2$}, \qquad \qquad
{\{}\overline{Q}{}^a, Q_b{\}} = \lambda \left[
(L^2){}^a{}_b - L{}^a{}_b\big(\langle L \rangle - \alpha\big)
-\textstyle{\frac 12}\delta{}^a{}_b\big(\langle L^2 \rangle \!-\! \langle L \rangle^2 \!+\! (n\!-\!1\!+2\alpha)\langle L \rangle \big) 
+ {\texttt c}\delta^a{}_b \right].
$$
\end{widetext}
The equivalence of $sl_{(1)}(n/1)$ to $sl(n/1)\cong A(n\!-\!1,0)$ can be seen as follows. Firstly, by a suitable rescaling of the odd generators, the overall constant $\lambda$ can be set to 1.
Then by redefining the $L^a{}_b$ generators via $L^a{}_b \rightarrow L^a{}_b - \delta^a{}_b \texttt{c} /(n\!-\!1)$, and finally shifting to $K^a{}_b := L^a{}_b - \delta^a{}_b \langle L \rangle/n$, $Z= \langle L \rangle/(n\!-\!1)$, we have the standard presentation in terms of the even subalgebra $sl(n) \times gl(1)$, ${\{}\overline{Q}{}^a,{Q}_b{\}} = K^a{}_b - Z \, \delta^a{}_b$, wherein the central charge parameter \texttt{c} no longer appears explicitly. Similar rescalings and parameter adjustments can be enacted in the remaining cases, but for explicit purposes we retain the somewhat redundant parametrisations as above.
The proof that these generalised structure constants fulfil the Jacobi identities is straightforward. The mixed even and odd cases
are satisfied automatically, and the only check needed is that ${[} {\{}\overline{Q}{}^a,{Q}_b{\}}, Q_c {]}$ is antisymmetric in $b$ and $c$. For example, in the $p=1$ case, there is no condition on \texttt{c}, and
${[} L^a{}_b - \delta^a{}_b \langle L \rangle, Q_c{]} = - \delta^a{}_c Q_b + \delta^a{}_b Q_c$, as required (because ${[} \langle L \rangle, Q_c{]} = - Q_c$). Surprisingly, while the resulting system of homogeneous equations allows parametrised solutions for the quadratic and cubic deformed cases $p=2,3$, it turns out that these equations have full rank for additional quartic coefficients -- there exists no quartic polynomial deformation. \hfill $\diamondsuit$
\\[.2cm]

We now recall some facts about representations of the Lie algebra $gl(n)$. As is well known, highest weight representations are labelled by $n$ invariants $(\lambda_1,
\lambda_2, \cdots, \lambda_n)$. Acting in such representations, the  matrix array $L^a{}_b$ fulfils\cite{BrackenGreen1971vo,green1971ci} an analogue of the Cayley-Hamilton identity, a degree $n$ polynomial relation in powers of $L$ of the form $\prod_{r=1}^n (L-\alpha_r) = 0$, with roots $\alpha_r := \lambda_r+n-r$. Moreover, in the case of coincident roots, there is a minimal identity of the appropriately reduced degree. These statements still generically apply in infinite dimensional highest-weight representations (see for example \cite{Gould1984}). 

The existence of zero step atypical representations of polynomial $gl(n)$ superalgebras \cite{JarvisRudolphYates2011} follows from the simple observation that in certain cases, for a given module of the even subalgebra, the polynomials ${\mathcal P}_{(p)}(L)$ can be brought into exact correspondence with the appropriate polynomial identity for that representation -- the right-hand side (\ref{eq:CurlyPdefs})  of the anticommutation relation ${\{} Q_a,\overline{Q}{}^b{\}}$ vanishes, and so the odd generators can be set to zero; the supersymmetry is carried entirely on a single multiplet of the even subalgebra.


\begin{table}
\caption{\label{tab:Polys}The polynomials ${\mathcal P}_{(p)}(L)$ for $p=2,3$. Monomials
$k(\kappa)\equiv L^k\langle L^\kappa\rangle$ are listed with their accompanying coefficients, where $\kappa = (\kappa_1,\kappa_2,\cdots)$ and
$\langle L^\kappa\rangle \equiv \langle L^{\kappa_1} \rangle \langle L^{\kappa_2} \rangle\cdots$. For example, the third row entries for ${\mathcal P}_{(3)}$ should be read as the term $-\frac 12L^a{}_b(\langle L^2 \rangle \!-\! \langle L \rangle^2 \!+\! (2\alpha\!+\!n\!-\!1)\langle L \rangle \!-\!2\beta)$ in ${\{}\overline{Q}{}^a, Q_b{\}}$. 
}
\begin{ruledtabular}
\begin{tabular}{llllll}
$2(0)$ &&& \\
$1$&&& \\[.2cm]
\hline
$1(1)$ & $1(0)$ &&  \\
$-1$&$\alpha$&& \\[.2cm]
\hline
$0(2)$& $0(1^2)$& $0(1)$&$0(0)$  \\
$-\frac 12$&$\frac 12$&$\!-\!\alpha\!-\!\frac 12(n\!-\!1)$&\texttt{c} \\[.2cm]
\hline
\hline
$3(0)$ &&& \\
$1$&&&\\[.2cm]
\hline
$2(1)$ & $2(0)$ &&  \\
$-1$&$\alpha$&& \\[.2cm]
\hline
$1(2)$& $1(1^2)$& $1(1)$& $1(0)$ \\
$-\frac 12$&$\frac 12$&$\!-\!\alpha\!-\!\frac 12(n\!-\!1)$&$\beta$ \\[.2cm]
\hline
$0(3)$& $0(21)$& $0(1^3)$&\\
$-\frac 13$&$\frac 12$&$-\frac 16$&\\[.2cm] 
\hline
$0(2)$& $0(1^2)$& $0(1)$ &$0(0)$  \\
$\!-\!\frac 12\alpha\!-\!\frac 16(2n\!-\!3)$&$\frac 12\alpha\!+\!\frac 16(3n\!-\!4)$&
$\!-\!\beta\!-\!\frac 12(n\!-\!1)\alpha$&\texttt{c} \\
&&$\!-\!\frac 16(2n^2\!\!-\!3n\!+\!2)$&\\[.2cm]
\end{tabular}
\end{ruledtabular}
\end{table}

\section{Conformal supermultiplets without superpartners}

It is well known that the Lie algebra of the spacetime conformal group is isomorphic to $su(2,2)$,
a real form of $sl(4)$. Equivalents of the above polynomial superalgebras exist for $su(2,2)$. In this non-compact case, the 
requirement of having unitary representations imposes certain hermiticity conditions on the Gel'fand generators (and matching  conditions on the odd generators (supercharges) of the putative superalgebras). Calling them $J^a{}_b$, $Q_a$, and $\overline{Q}{}^b$ and introducing the pseudo-unitary metric
$\eta = \mbox{diag}(1,1,-1,-1)$, $a,b = 1,2,3,4$, we have 
\begin{equation}
\label{eq:Hermiticity}
(J^a{}_b)^\dagger = \eta^{b}{}_{b'} \, J^{b'}{}_{a'} \, \eta^{a'}{}_a, \qquad  (Q_a)^\dagger = \eta^{a}{}_{b'}\overline{Q}{}^{b'}.
\end{equation}
In this real basis, the commutation relations and polynomial deformations are still the same as those given above for $L^a{}_b$. Most importantly, 
the polynomial characteristic identity continues to hold in suitable (infinite dimensional) representations.

The classification of the physical, positive energy field multiplets of conformal symmetry is as follows\cite{Mack1969}. These are technically lowest weight 
unitary irreducible representations labelled $(j_1,j_2;d)$, where $(j_1,j_2)$ is the spin, and $d$ the conformal dimension. Of the five classes\cite{Mack1969} of such unitary representations, the one of interest is the massless unirreps, with $j_1j_2=0$, and $d=j_1+j_2+1$, namely either $(j_1,0; j_1+1)$, $(0,j_2; j_2+1)$ or $(0,0;1)$.  From above therefore, the $4\! \times \!4$ array $J$, when acting on such multiplets, will generically satisfy a polynomial identity of at most \emph{quartic} degree. In the Appendix,
\S A, these conformal multiplets are constructed within the Fock space of two pairs of bosonic creation and annihilation operators, and the conformal generators are explicitly shown to satisfy a \emph{quadratic} characteristic identity\cite{BarutBohm1969rcr}.

Zero-step representations for the quadratic case $su_{(2)}(2,2/1)$ are produced as follows. Referring to (\ref{eq:CurlyPdefs}) we set $\lambda=1$  (and $n=4$) and consider the quadratic polynomials ${\mathcal P}_{(2)}(J;\alpha,\texttt{c})$: \\[.2cm]

\noindent
\emph{Theorem. Zero-step supermultiplets for $su_{(2)}(2,2/1)$}.\hfill \\ 
For the choice $\alpha=-3$, ${\texttt c}=0$ of central charge parameters, the deformed anticommutator for the supercharges for the quadratic superalgebra $su_{(2)}(2,2/1)$ is
$$
{\mathcal P}_{(2)}(J; -3,0) = J^2 -J\big(\langle J\rangle \!+\!3\big) 
-\textstyle{\frac 12}\big(\langle J^2\rangle -\langle J\rangle(\langle J\rangle \!+\!3) \,  \big).
$$
For the massless field multiplets of the conformal group $SU(2,2)$ (appendix, \S A) the generators satisfy the quadratic identity
$J^2 = J(\langle J\rangle + 3)$ with $\langle J\rangle + 3$ given by  $2j_1\!+\!1$ or $-\!2j_2\!+\!1$ for $(j_1,0; j_1\!+\!1)$ and $(0,j_2; j_2\!+\!1)$, respectively. From above, ${\mathcal P}_{(2)}(J; -3,0)$ is identically zero, and so the massless conformal field multiplets  form zero-step supermultiplets for the quadratic superalgebra $su_{(2)}(2,2/1)$. \hfill $\diamondsuit$\\[.2cm]

In other words, it is possible to fulfil the deformed anticommutation relations (\ref{eq:CurlyPdefs}) of $su_{(2)}(2,2/1)$  if the supercharge generators $Q_a$, $\overline{Q}{}^b$ have \emph{identically zero} action on the underlying fields: in each case the supersymmetry is carried \emph{entirely on a single multiplet of the even subalgebra}.

\section{Discussion}

The present exploration of alternative forms of spacetime supersymmetry mirrors work to investigate potential violations of ordinary relativistic invariance from physics at the highest (unification- or Planck- scale) energies. In that context, the occurrence of a fundamental energy scale engenders possible experimental signals of new physics beyond certain threshholds \cite{JacobsonLiberatiMattingly2003}. This corresponds mathematically to certain small correction terms in quantities such as dispersion relations, which become significant for such processes. The situation with $su_{(2)}(2,2/1)$ is technically similar -- the admixture of nonlinear quadratic deformation terms in the algebraic relations (\ref{eq:CurlyPdefs}), relative to the standard superalgebra, is governed by the parameter ratio $\lambda:\lambda \alpha$ (note that in the limit $\lambda \rightarrow 0$, with $\lambda \alpha= 1$, we have $su_{(2)}(2,2/1) \rightarrow su(2,2/1)$).
However, in this case the very presence of the additional quadratic terms, however suppressed, guarantees the existence of zero-step supermultiplets.
In a scenario where present, low energy (massless) fields are descended from more fundamental entities at unification scales, the cases of interest are 
$(\frac 12,0;\frac 32)$, $(0,\frac 12;\frac 32)$, $(1,0;2)$, $(0,1;2)$ and $(0,0;1)$ (spin-$\frac 12$ fermions and spin-$1$ vector bosons of either chirality, and scalar, respectively). This suggests that the fundamental leptons, quarks, massless vectors and Higgs scalars of the standard model are not accompanied by superpartners. Note however that ``massive'' spin-1 gauge fields may not be so privileged if they derive from vector multiplets such as $(\textstyle{\frac 12},\textstyle{\frac 12};2)$.

The concern of this work has been technical, rather than constructive as to how the proposed deformed superalgebras might arise in particle models going beyond local relativistic quantum field theory \cite{HaagLopuszanskiSohnius1975}. We have been at pains simply to establish one explicit example of supersymmetry without superpartners -- the question of similar mechanisms for extended analogues of Poincar\'{e} or related spacetime supersymmetries is open. Hints for model-building are provided however by early attempts to generalise strong interaction current algebras to incorporate both baryon (trilinear) and meson (bilinear) fields \cite{DelbourgoSalamStrathdee1966mfr}. Recently, motivated by an analysis of algebraic structures in Hamiltonian lattice gauge theories, a variety of polynomial $gl(n)$ superalgebras was identified \cite{JarvisRudolph2003} via concrete free field constructions, including quadratic superalgebras as a case in point \cite{JarvisRudolphYates2011}.  In the present context, trilinear fermionic (technicolour?) gauge invariant fields could indeed provide the ingredients for the Noether supercharges of quadratic superalgebras.  Based on this, it might be conjectured then that the notion of `secret' supersymmetry -- supersymmetry without superpartners -- is suggestive of further particle substructure at fundamental scales.

\begin{acknowledgments}
PDJ thanks the Department of Theoretical Physics, University of Leipzig for hospitality during collaborative visits, and the Alexander von Humboldt Foundation for support. Similar appreciation is expressed to the Australian-American Fulbright Foundation, and staff and colleagues at the Department of Statistics, University of California Berkeley, as well as the Department of Physics, University of Texas Austin, for visits as an Australian senior Fulbright scholar, during part of this work. Discussions with A. J. Bracken, R. Delbourgo, G. Rudolph, and R. B. Zhang are gratefully acknowledged.
 
 \end{acknowledgments}

\appendix

\section{Fock space construction of massless conformal multiplets}

Unitary irreducible representations of $su(2,2)$ of the massless type considered in this letter can be constructed explicitly as ladder subspaces inside the Fock space of two pairs of bosonic creation and annihilation operators (each of which realizes Schwinger's original $SU(2)$ model). These operators, $a_i$, $b_j$, $i,j = 1,2$, and their conjugates, satisfy the usual bosonic oscillator algebra
\begin{equation}
{[}a_i,a_i{}^\dagger{]} = \delta_{i,j} = {[}b_i,b_j{}^\dagger{]},
\end{equation}
with $\widehat{N}_a = \sum_i a_i{}^\dagger a_i$, $\widehat{N}_b = \sum_j b_j{}^\dagger b_j$ being the number operators.
Consider the four-component objects $a_1$, $a_2$, $a_3 \equiv b_1{}^\dagger$, $a_4 \equiv b_2{}^\dagger$ and 
$a^1 := a_1{}^\dagger$, $a^2 := a_2{}^\dagger$, $a^3 := -b_1$, $a^4 := -b_2$. 
Then for $a,b = 1,\cdots,4$ define $J^a{}_b = a^a a_b$ which satisfy (\ref{eq:GLnGelfandBasis}) and (\ref{eq:Hermiticity}). 
The quadratic characteristic identity $(J^2)^a{}_b =(\langle J \rangle +3) J^a{}_b $ is easily established, where the
linear Casimir is $\langle J \rangle = \widehat{N}_a - \widehat{N}_b -2$.
Within the Fock space spanned by the number states $(a_1{}^\dagger)^{n_1}(a_2{}^\dagger)^{n_2}(b_1{}^\dagger)^{n_3}(b_2{}^\dagger)^{n_4}|0,0,0,0\rangle$, subspaces with fixed eigenvalue $\Delta$ of  $\widehat{N}_a - \widehat{N}_b$ are invariant, and these are in turn built upon subspaces with eigenvalues $(N_a, N_b)  = (\Delta,0)$ or $(0,-\Delta)$ depending on the sign of $\Delta$. Given the conformal dimension $\widehat{d} =\textstyle{\frac 12} \sum_{a,b} \eta^a{}_b J^b{}_a= \textstyle{\frac 12}(J^1{}_1+  J^2{}_2- J^3{}_3-  J^4{}_4) =   \textstyle{\frac 12}(N_a+N_b)+1$, for $\Delta= 2j_1$, $\Delta=-2j_2$ these are the required massless multiplets $(j_1,0;j_1+1)$ and  $(0,j_2;j_2+1)$  respectively.


\end{document}